\RequirePackage[OT1]{fontenc}
\documentclass[conference]{IEEEtran}
\IEEEoverridecommandlockouts
\usepackage{cite}
\usepackage{amsmath,amssymb,amsfonts}
\usepackage{algorithmic}
\usepackage{graphicx}
\usepackage[caption=false,font=footnotesize]{subfig}
\usepackage{booktabs}
\usepackage{textcomp}
\usepackage{xcolor}
\usepackage{stfloats}
\usepackage{cuted}
\usepackage{flushend}
\usepackage{placeins}
\usepackage{capt-of}
\def\BibTeX{{\rm B\kern-.05em{\sc i\kern-.025em b}\kern-.08em
    T\kern-.1667em\lower.7ex\hbox{E}\kern-.125emX}}

\begin{document}

\title{Mutual Coupling and Visibility Effects in 2D Holographic Curvature-Reconfigurable Apertures\\
}

    
\author{%
\begin{tabular}{@{}c@{\hspace{3.0em}}c@{}}
\begin{minipage}{0.43\textwidth}\centering
Hang Lin\\
{\itshape School of Information Science and Electronic Engineering}\\
{\itshape Shanghai Jiao Tong University}\\
{Shanghai, China}\\
{linhang7346@sjtu.edu.cn}
\end{minipage}
&
\begin{minipage}{0.43\textwidth}\centering
Shu Sun\\
{\itshape School of Information Science and Electronic Engineering}\\
{\itshape Shanghai Jiao Tong University}\\
{Shanghai, China}\\
{shusun@sjtu.edu.cn}
\end{minipage}
\end{tabular}%
}

\maketitle

\begin{abstract}
Holographic multiple-input multiple-output arrays provide high spatial resolution and flexible wavefront control, but curved or conformal deployments jointly introduce curvature, mutual coupling (MC), and direction-dependent visibility. This paper studies a two-dimensional (2D) holographic curvature-reconfigurable aperture (HoloCuRA). A geometry-based model characterizes element positions, steering, and spatial correlation, while an impedance-based coupling matrix and a visibility vector capture electromagnetic interactions and shadowing. Relative directivity and spectral efficiency are evaluated versus curvature. Results show that MC mainly reshapes local directivity fluctuations, whereas visibility dominates the global pattern and spectral-efficiency variation under both isotropic scattering and CDL-A channels.
\end{abstract}


\section{Introduction}
Holographic multiple-input multiple-output (HMIMO) uses densely spaced elements to approximate a quasi-continuous electromagnetic aperture, enabling high spatial resolution and flexible wavefront control for sixth-generation (6G) systems \cite{huang2020holographic}. Its analysis requires joint treatment of spatial degrees of freedom, aperture geometry, and propagation, including near-field effects for large apertures \cite{pizzo2022fourier,10819602}.

Most studies assume planar apertures \cite{deng2022reconfigurable,11513410}, whereas curved or conformal deployments change element positions, steering, correlation, and projected aperture. Dense spacing also introduces mutual coupling (MC), affecting excitation, efficiency, and channel eigenvalues \cite{11719648,yuan2023effects}; coupling-matrix models are thus important for compact and superdirective arrays \cite{han2022coupling,pizzo2024superdirectivity}.

Direction-dependent visibility further limits the participating aperture. We study a two-dimensional holographic curvature-reconfigurable aperture (2D HoloCuRA) by combining curvature-dependent geometry and correlation with impedance-based MC and geometric visibility. Relative directivity and spectral efficiency distinguish the local impact of MC from the global influence of visibility.

\section{System Model}
\begin{figure}[t]
    \centering
    \includegraphics[width=0.55\columnwidth]{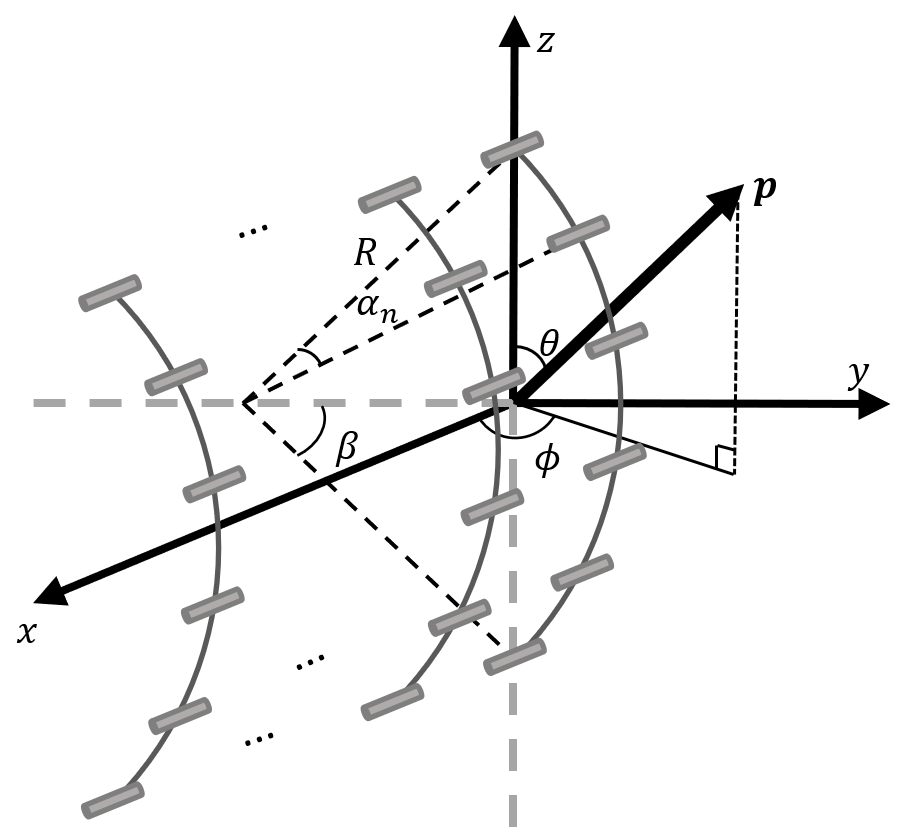}
    \caption{2D HoloCuRA in the coordinate system.}
    \label{Fig2DHoloCuRAwithCoordinateSystem}
\end{figure}

Consider the 2D HoloCuRA in Fig.~\ref{Fig2DHoloCuRAwithCoordinateSystem}, oriented toward the positive $y$-axis for flexible directional coverage.
The array contains $M\times N$ elements, with $N$ along the curved direction and $M$ along $x$. The $(m,n)$th element is located at \cite{11174411}
\begin{equation}
\boldsymbol r_{mn}=\begin{bmatrix}(m-1)d\\R\cos(\beta-\alpha_n)-R\cos\beta\\R\sin(\beta-\alpha_n)\end{bmatrix},\quad
\substack{m=1,\ldots,M\\n=1,\ldots,N},
\label{eq:position}
\end{equation}
where $\alpha_n=(n-1)L/[(N-1)R]$, $R=L/(2\beta)$, and $L$ is the arc length. With dipole length $l$ and gap $\delta$, $d=l+\delta$ and the aperture is $D\times L$, where $D=Ml+(M-1)\delta$. All dipoles remain parallel to $x$, which lies in every local tangent plane. The planar case uses the $\beta\to0$ limit.

For direction $(\phi,\theta)$, define the unit direction vector $\hat{\boldsymbol p}=(\sin\theta\cos\phi,\sin\theta\sin\phi,\cos\theta)^{\rm T}$ and wavenumber $k=2\pi/\lambda$. Steering and excitation are $\boldsymbol a_0=[e^{jk\hat{\boldsymbol p}\cdot\boldsymbol r_{11}},\ldots,e^{jk\hat{\boldsymbol p}\cdot\boldsymbol r_{MN}}]^{\rm T}$ and $\boldsymbol j_0=[j_{11},\ldots,j_{MN}]^{\rm T}$. The relative maximum directivity is \cite{pizzo2024superdirectivity}
\begin{equation}
D_0^{\rm rel}(\phi,\theta)=\frac{1}{MN}\boldsymbol a_0^{\rm T}\mathbf R_0^{-1}\boldsymbol a_0^*,\quad
\mathbf R_0=\iint s\,\boldsymbol a\boldsymbol a^{\rm H}\,d\phi d\theta,
\label{eq:d0}
\end{equation}
where $\mathbf R_0$ is the spatial correlation matrix and $s(\phi,\theta)$ is the normalized spatial scattering function. Under isotropic scattering \cite{11719648},
\begin{equation}
[\mathbf R_0]_{(m_1,n_1),(m_2,n_2)}=\operatorname{sinc}\!\left(\frac{2\|\boldsymbol r_{m_1n_1}-\boldsymbol r_{m_2n_2}\|_2}{\lambda}\right).
\end{equation}

For compact arrays, MC modifies the response as \cite{han2022coupling}
\begin{equation}
J_c=\boldsymbol a_0^{\rm T}\mathbf C\boldsymbol j_0,\quad
D_c^{\rm rel}=\frac{1}{MN}\boldsymbol a_0^{\rm T}\mathbf C\mathbf R_0^{-1}\mathbf C^{\rm H}\boldsymbol a_0^*,
\label{eq:mcdir}
\end{equation}
where $\mathbf C\in\mathbb C^{MN\times MN}$ is the coupling matrix.

We idealize the backing as suppressing rear-side response. With $\psi_n=\beta-\alpha_n$, the local front normal is $\boldsymbol n_n=(0,\cos\psi_n,\sin\psi_n)^{\rm T}$. The local half-space visibility rule is
\begin{equation}
v_{mn}=\mathbf1\{\boldsymbol n_n^{\rm T}\hat{\boldsymbol p}\geq0\},\quad
\boldsymbol a_v=\boldsymbol v\odot\boldsymbol a_0,
\label{eq:visibility}
\end{equation}
where $\hat{\boldsymbol p}$ points toward the departure direction or arrival source. For directivity, replace $\boldsymbol a_0$ by $\boldsymbol a_v$ in \eqref{eq:d0} and \eqref{eq:mcdir}, keeping $\mathbf R_0$ as the reference correlation matrix.

\section{Numerical Results and Analysis}
Table~\ref{tab:simulation_parameters} lists the aperture parameters. Since bending changes positions but not the parallel dipole axes, we use the classical parallel-dipole mutual impedance $Z_{(m_1,n_1),(m_2,n_2)}=R+jX$ \cite{balanis2016antenna}:
\begin{align}
R={}&\frac{\eta}{8\pi}[-\cos w_0 A_C+\sin w_0 A_S],\notag\\
X={}&\frac{\eta}{8\pi}[-\cos w_0 B_S+\sin w_0 B_C],\label{eq:mutualZ}\\[-1mm]
A_C={}&-2C_i(w_1)-2C_i(w'_1)+C_i(w_2)+C_i(w'_2)\notag\\[-1mm]
&+C_i(w_3)+C_i(w'_3),\notag\\[-1mm]
A_S={}&2S_i(w_1)-2S_i(w'_1)-S_i(w_2)+S_i(w'_2)\notag\\[-1mm]
&-S_i(w_3)+S_i(w'_3),\notag\\[-1mm]
B_S={}&2S_i(w_1)+2S_i(w'_1)-S_i(w_2)-S_i(w'_2)\notag\\[-1mm]
&-S_i(w_3)-S_i(w'_3),\notag\\[-1mm]
B_C={}&2C_i(w_1)-2C_i(w'_1)-C_i(w_2)+C_i(w'_2)\notag\\[-1mm]
&-C_i(w_3)+C_i(w'_3),\notag\\[-1mm]
w_0={}&k\Delta_h,\notag\\[-1mm]
w_1={}&k(\sqrt{\Delta_d^2+\Delta_h^2}+\Delta_h),\quad
w'_1=k(\sqrt{\Delta_d^2+\Delta_h^2}-\Delta_h),\notag\\[-1mm]
w_2={}&k[\sqrt{\Delta_d^2+(\Delta_h-l)^2}+\Delta_h-l],\notag\\[-1mm]
w'_2={}&k[\sqrt{\Delta_d^2+(\Delta_h-l)^2}-\Delta_h+l],\notag\\[-1mm]
w_3={}&k[\sqrt{\Delta_d^2+(\Delta_h+l)^2}+\Delta_h+l],\notag\\[-1mm]
w'_3={}&k[\sqrt{\Delta_d^2+(\Delta_h+l)^2}-\Delta_h-l],\notag
\end{align}
where $\Delta_h=|m_1-m_2|d$ is the axial separation and $\Delta_d=2R|\sin[(\alpha_{n_1}-\alpha_{n_2})/2]|$ is the transverse chord distance. Curvature enters through these distances, without an orientation correction.
The impedance-to-coupling transformations at the transmitter (TX) and receiver (RX) are \cite{1167256}\par
{\footnotesize
\begin{equation}
\mathbf C_T=(1+z_S/z_A)\mathbf Z(\mathbf Z+z_S\mathbf I_{MN})^{-1},\;
\mathbf C_R=(z_A+z_L)(\mathbf Z+z_L\mathbf I_{MN})^{-1},
\label{eq:couplingTR}
\end{equation}}
where $\mathbf Z$ has self-impedance $z_A$ on its diagonal. Source and load impedances use conjugate matching, $z_S=z_L=z_A^*$.

\begin{table}[t]
\centering
\caption{Simulation Parameters}
\label{tab:simulation_parameters}
\footnotesize
\renewcommand{\arraystretch}{0.96}
\begin{tabular*}{0.97\columnwidth}{@{\extracolsep{\fill}}lc}
\toprule
Parameter & Value\\
\midrule
Carrier frequency $f$ & $30$ GHz\\
Elements $M\times N$ & $8\times 8$\\
Arc length $L$ & $0.0175$ m\\
Dipole length $l$ & $0.005$ m\\
Element gap $\delta$ & $0.0005$ m\\
\bottomrule
\end{tabular*}
\end{table}

\begin{table}[t]
\centering
\caption{Channel Simulation Parameters}
\label{tab:channel_simulation_parameters}
\footnotesize
\renewcommand{\arraystretch}{0.96}
\begin{tabular*}{0.97\columnwidth}{@{\extracolsep{\fill}}lc}
\toprule
Parameter & Value\\
\midrule
SNR & $10$ dB\\
Monte Carlo samples & $1000$\\
Isotropic rays $M_{\rm iso}$ & $50000$\\
CDL-A RMS delay spread & $100$ ns\\
CDL-A sampling rate & $15.36$ MHz\\
CDL-A maximum Doppler & $0$ Hz\\
\bottomrule
\end{tabular*}
\end{table}

\subsection{Directivity}
\begin{figure}[!ht]
\centering
\subfloat[]{\includegraphics[width=0.307\textwidth]{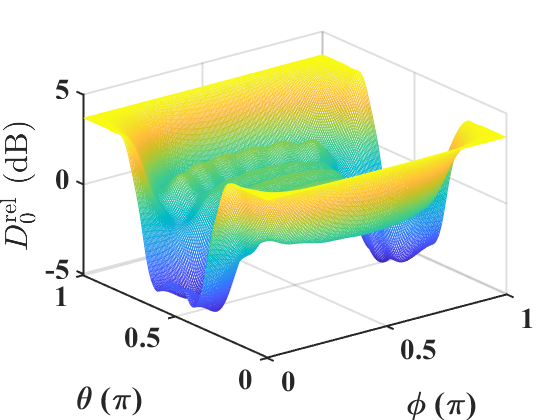}\label{fig:beta0U}}\hfill
\subfloat[]{\includegraphics[width=0.307\textwidth]{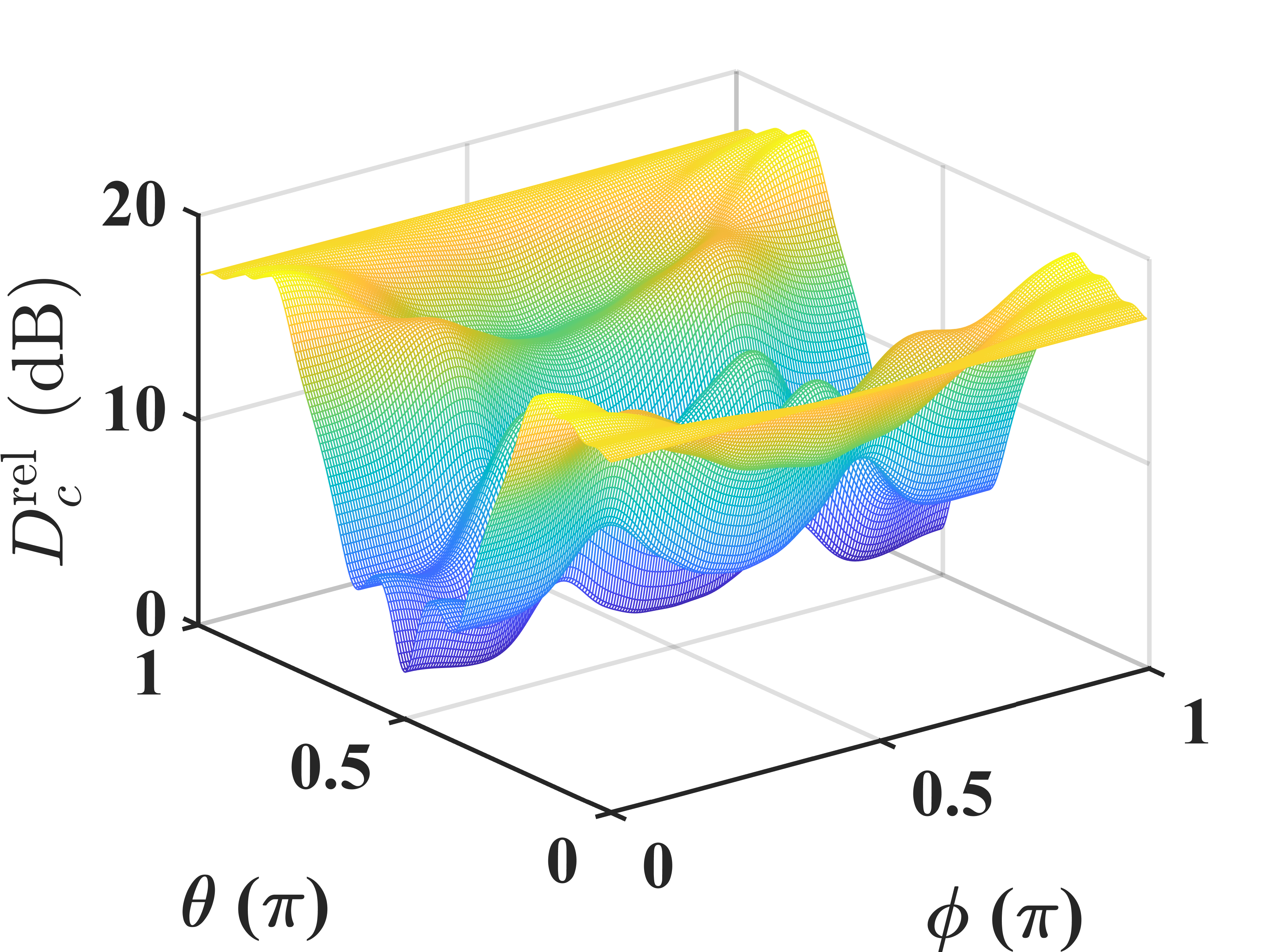}\label{fig:beta0C}}\hfill
\subfloat[]{\includegraphics[width=0.307\textwidth]{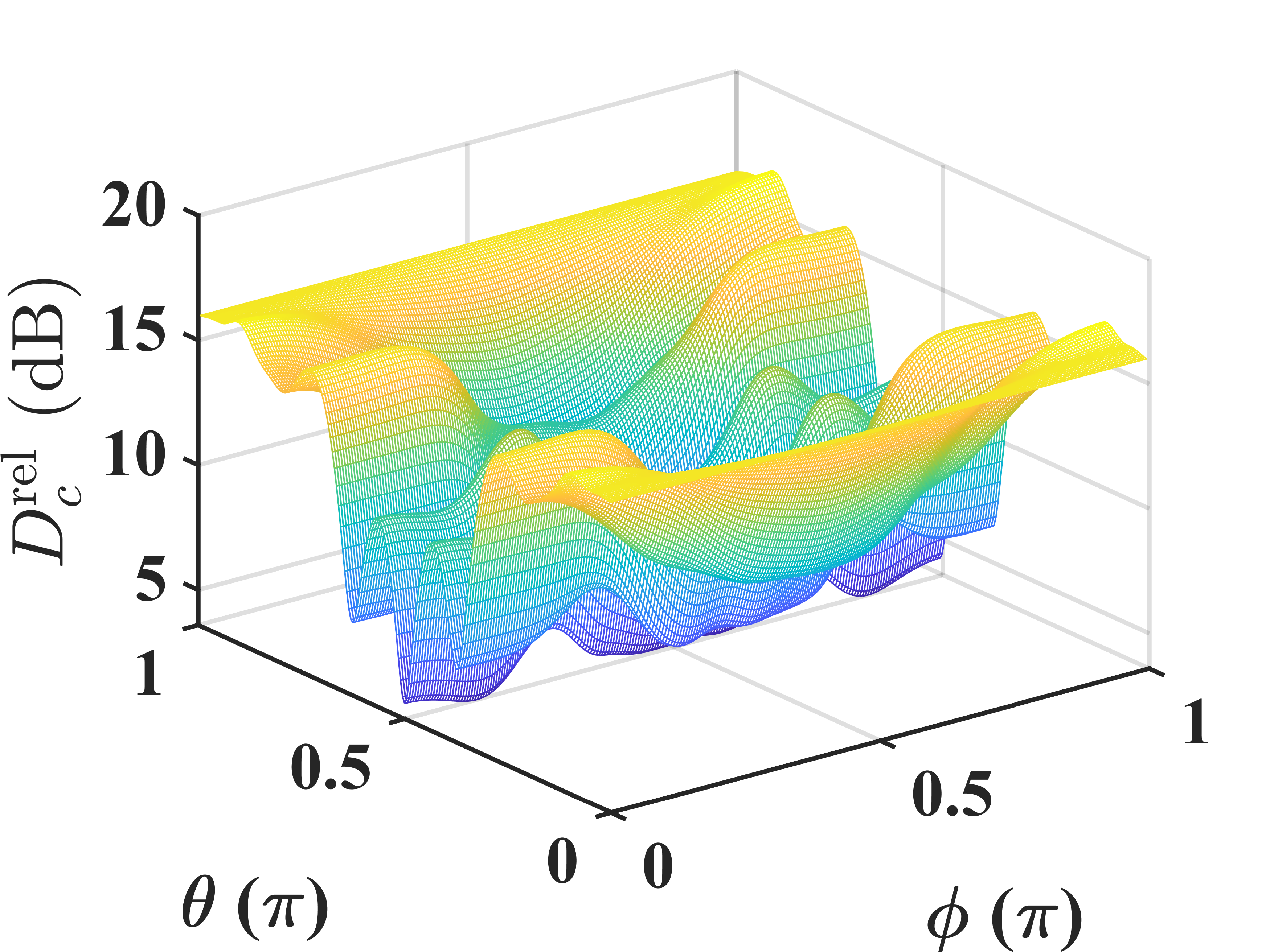}\label{fig:beta0F}}\\[-0.5mm]
\subfloat[]{\includegraphics[width=0.307\textwidth]{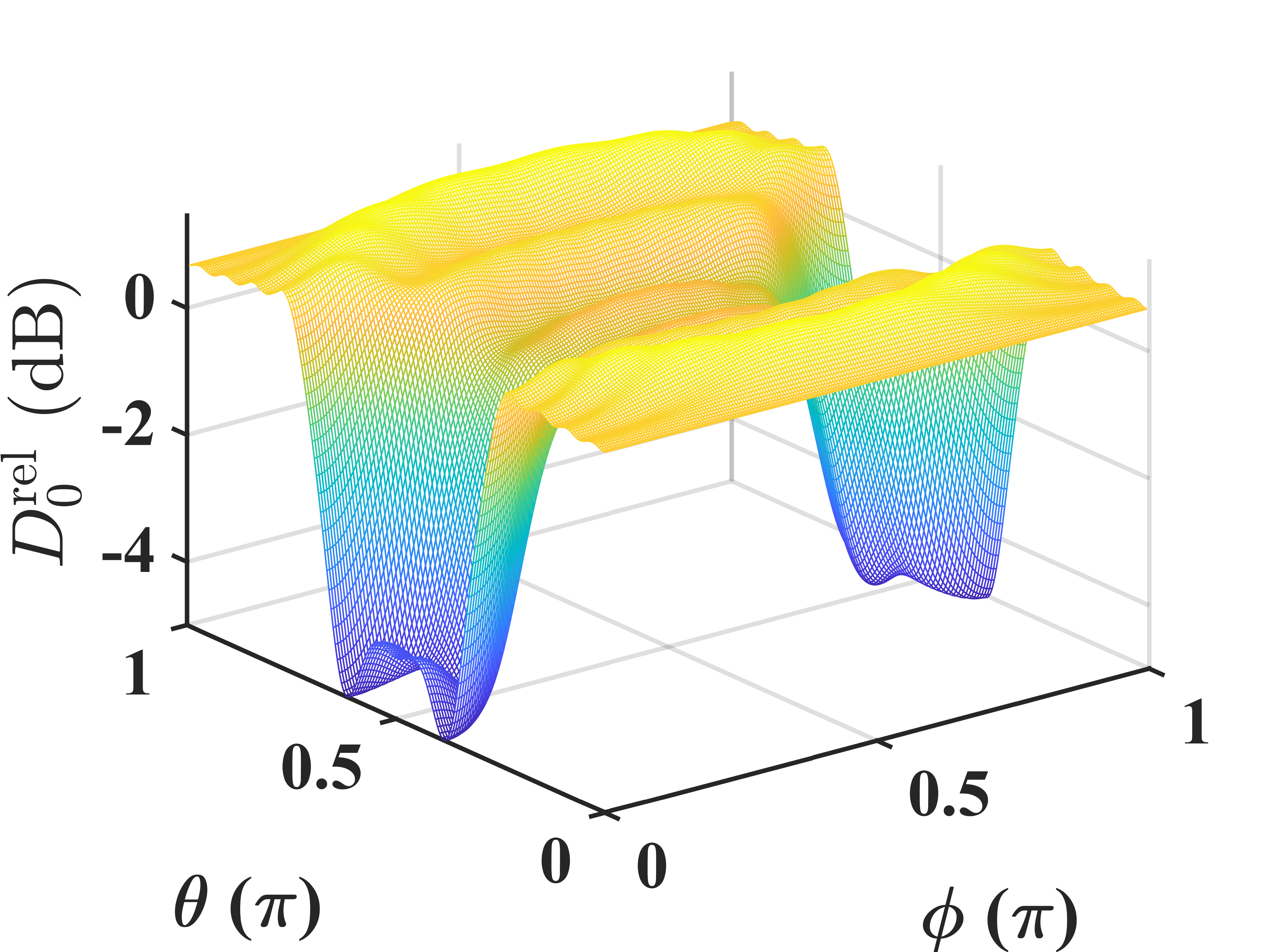}\label{fig:beta052U}}\hfill
\subfloat[]{\includegraphics[width=0.307\textwidth]{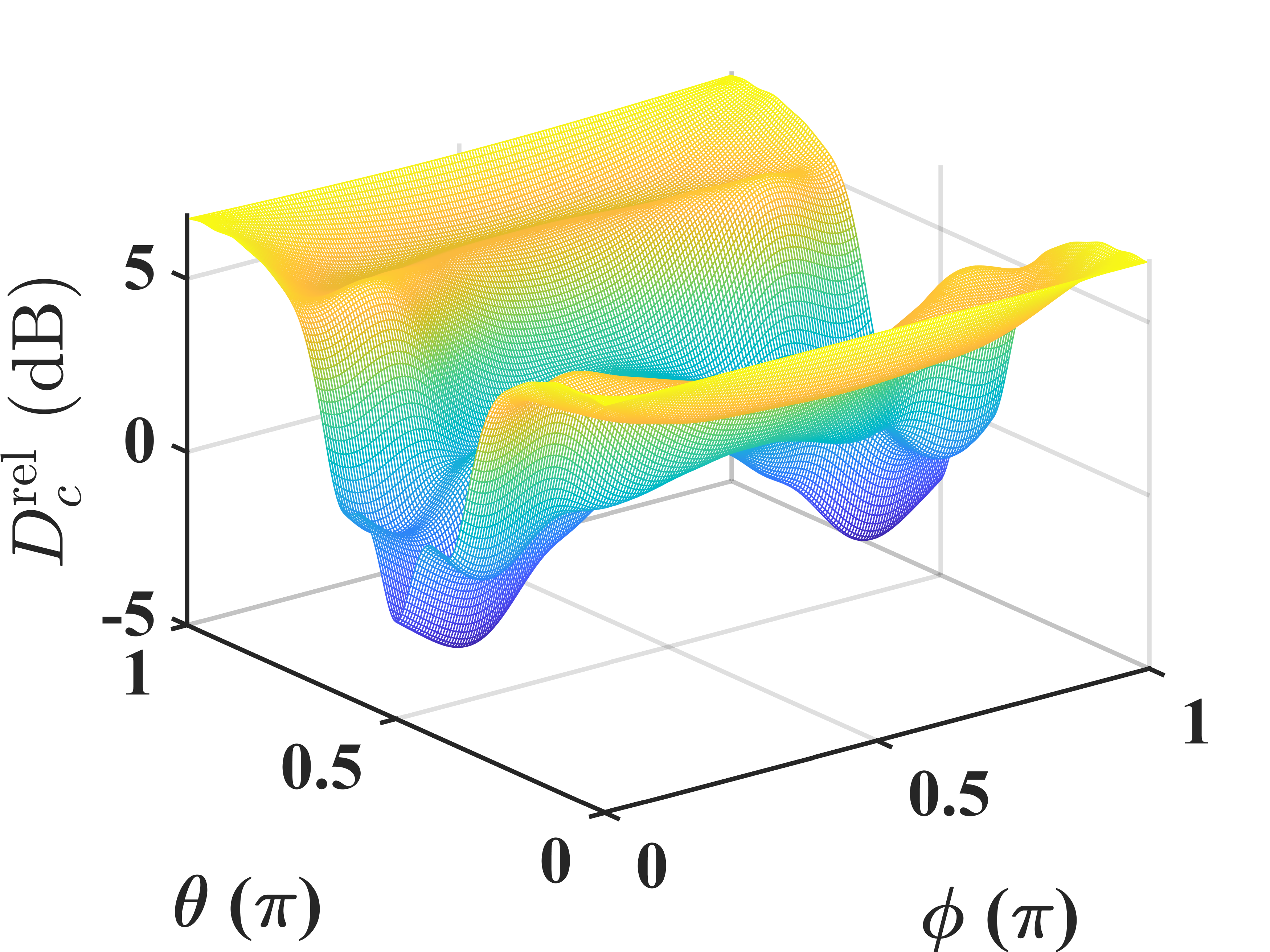}\label{fig:beta052C}}\hfill
\subfloat[]{\includegraphics[width=0.307\textwidth]{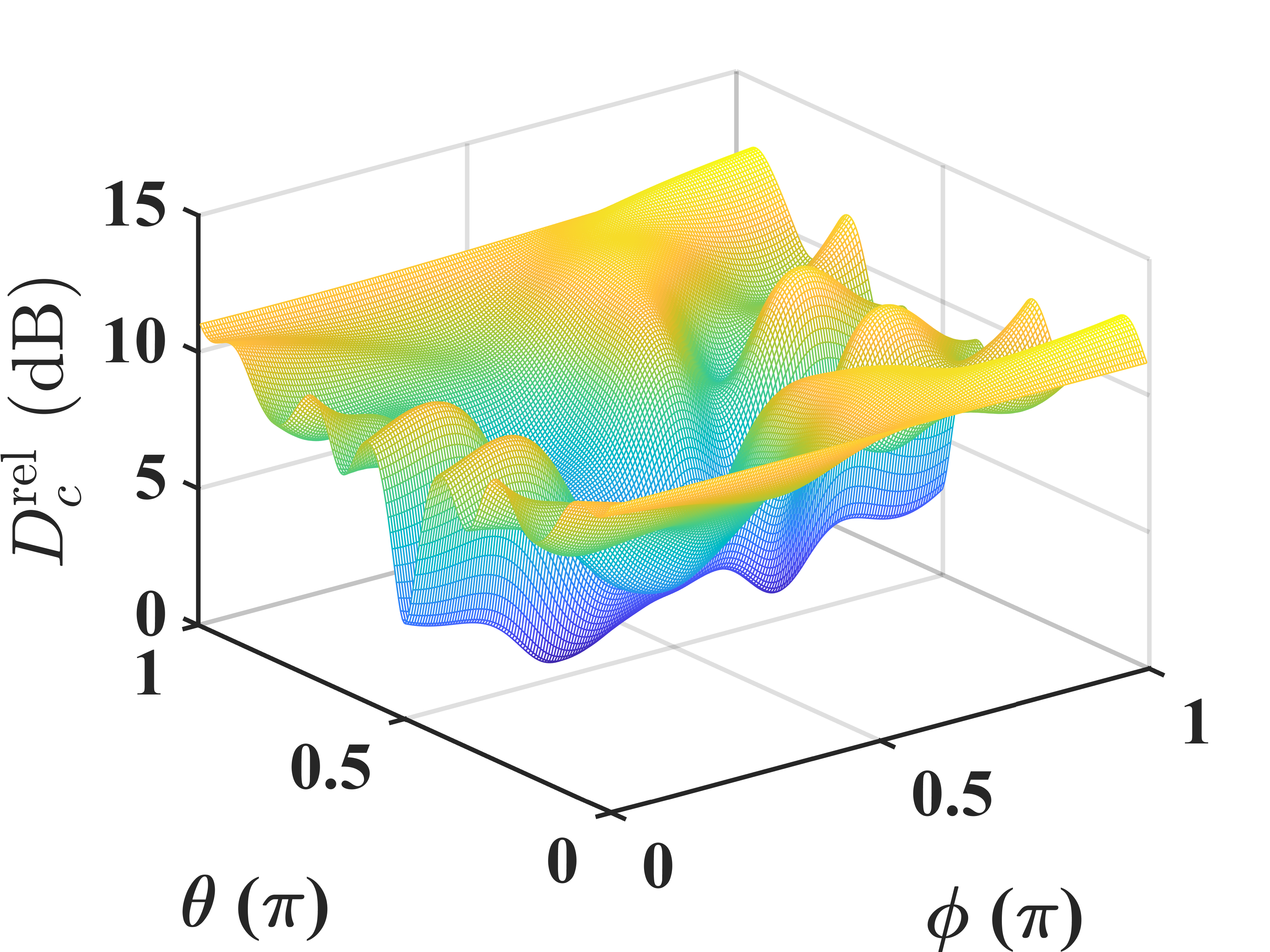}\label{fig:beta052F}}\\[-0.5mm]
\subfloat[]{\includegraphics[width=0.307\textwidth]{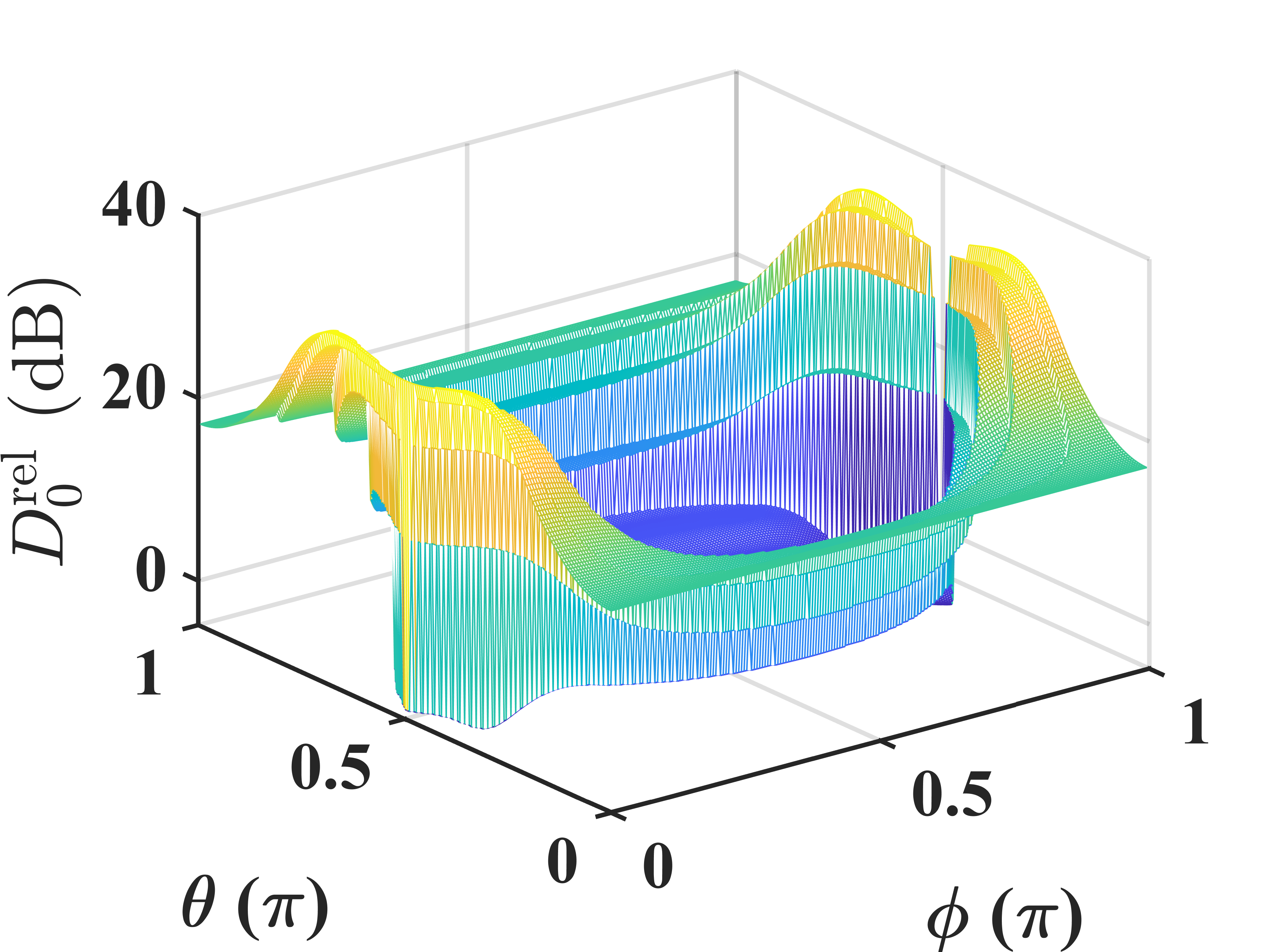}\label{fig:beta052UM}}\hfill
\subfloat[]{\includegraphics[width=0.307\textwidth]{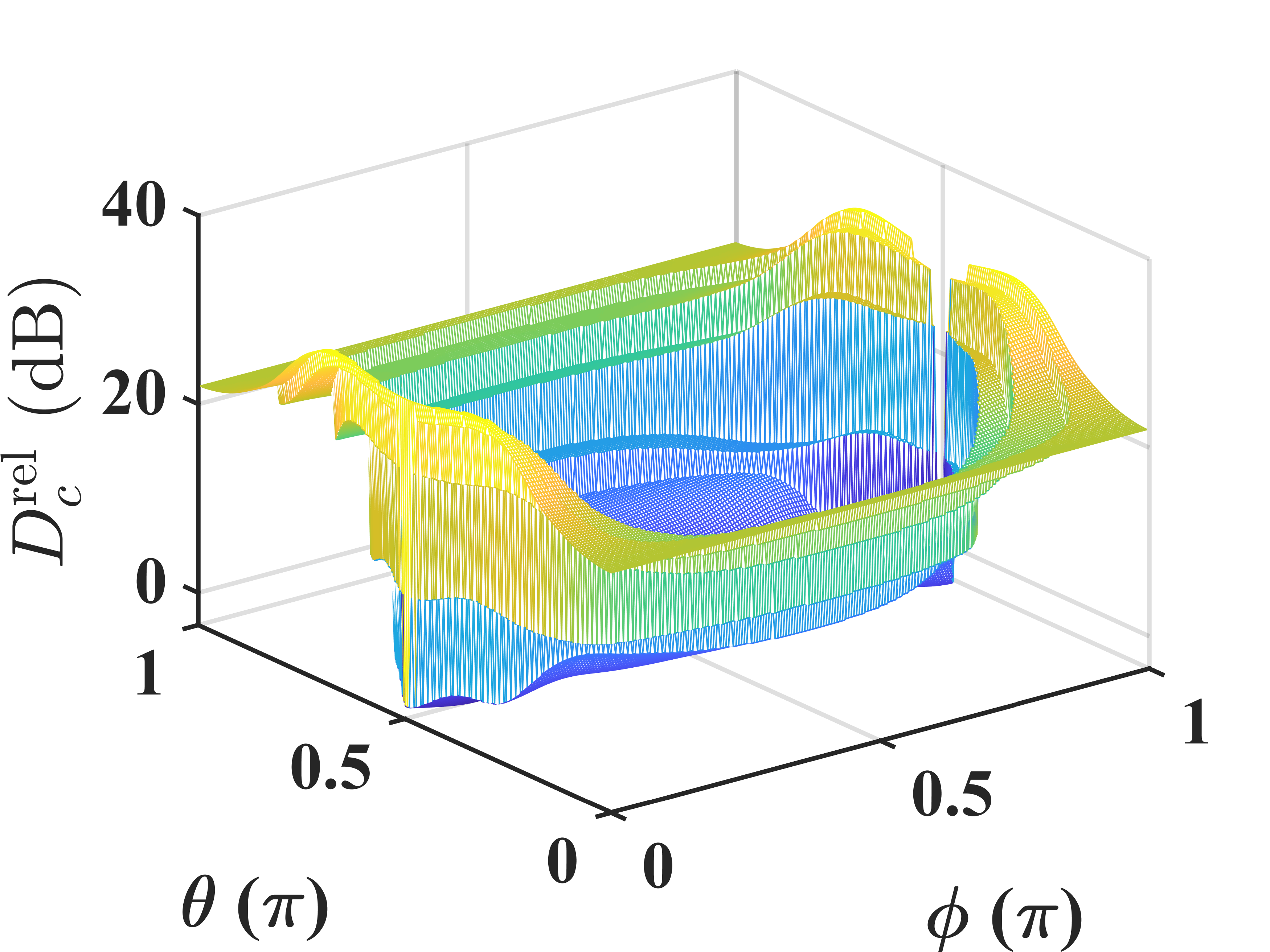}\label{fig:beta052CM}}\hfill
\subfloat[]{\includegraphics[width=0.307\textwidth]{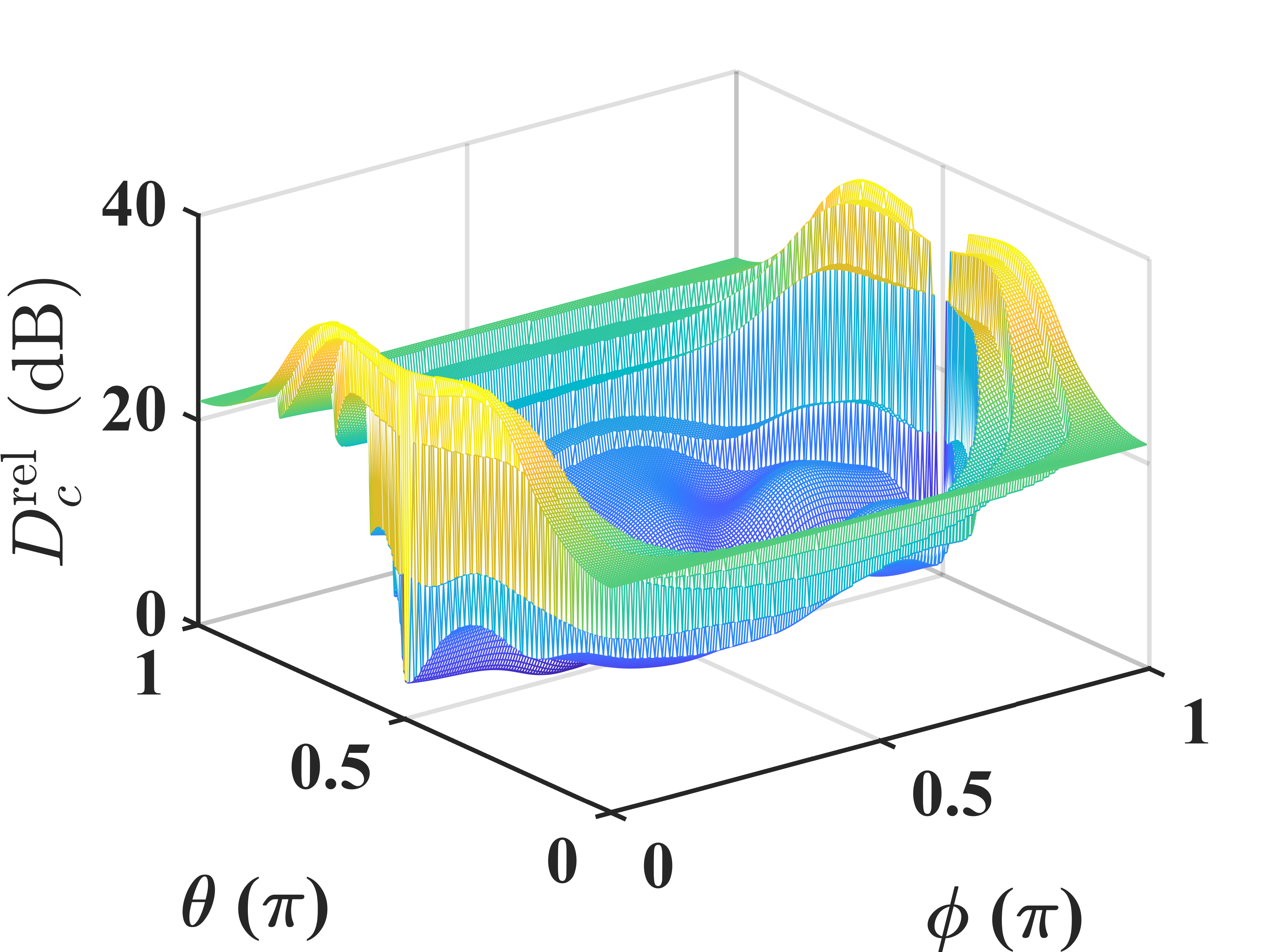}\label{fig:beta052FM}}
\caption{Relative directivity for $M=N=8$. Rows: $\beta=0$, $\beta=0.52$, and $\beta=0.52$ with visibility; columns: uncoupled, formula-based MC, and full-wave simulation.}
\label{Fig:RelDirec}
\end{figure}

Fig.~\ref{Fig:RelDirec} compares uncoupled, analytical-MC, and full-wave-impedance-based results for $\beta=0$ and $0.52$, with visibility added in the third row. The CST models retain the same $x$-directed dipoles. Their 64-port S parameters are converted to impedance matrices and then to coupling matrices via \eqref{eq:couplingTR}. Similar overall shapes and principal valleys support the analytical approximation for the studied geometry; local differences can reflect backing, ports, and edge effects. MC reshapes fluctuations, while visibility produces the largest global change by making the participating aperture direction-dependent.

For the planar aperture, coupling mainly changes the angular dynamic range. After bending, position-dependent phase shifts alter the locations and depths of the directivity valleys. Visibility additionally introduces angular boundaries at which participating columns change. Since the normal depends only on $n$, each switch affects an entire strip along $x$. This masking describes directional response rather than physical port disconnection; the full coupling matrix is retained. Local MC fluctuations can therefore coexist with a stronger global visibility effect.

\begin{figure}[!ht]
\centering
\subfloat[Uncoupled with visibility.]{\includegraphics[width=0.4\textwidth]{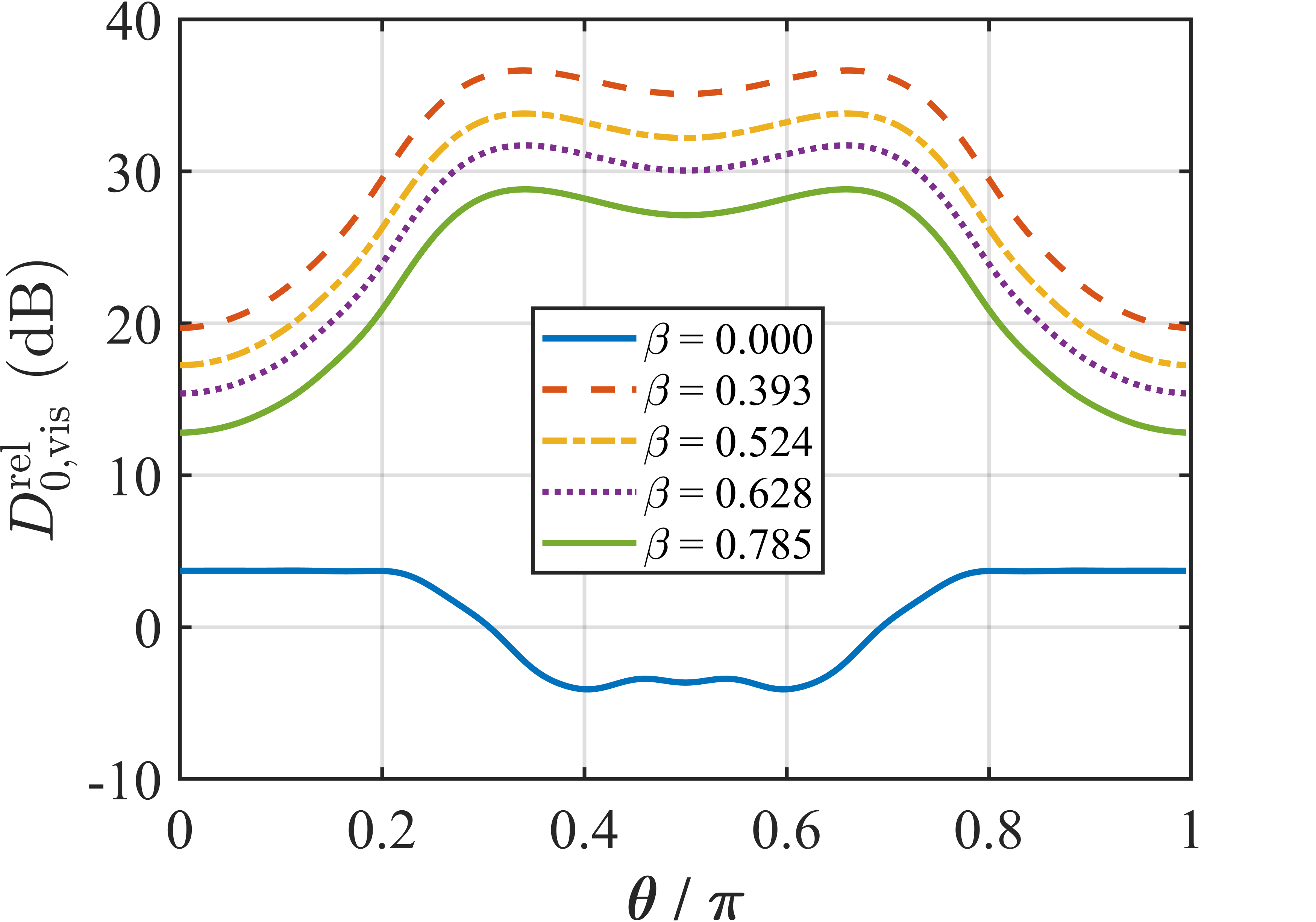}\label{fig:DirecFfpM}}
\hspace{0.02\textwidth}
\subfloat[Formula-based MC with visibility.]{\includegraphics[width=0.4\textwidth]{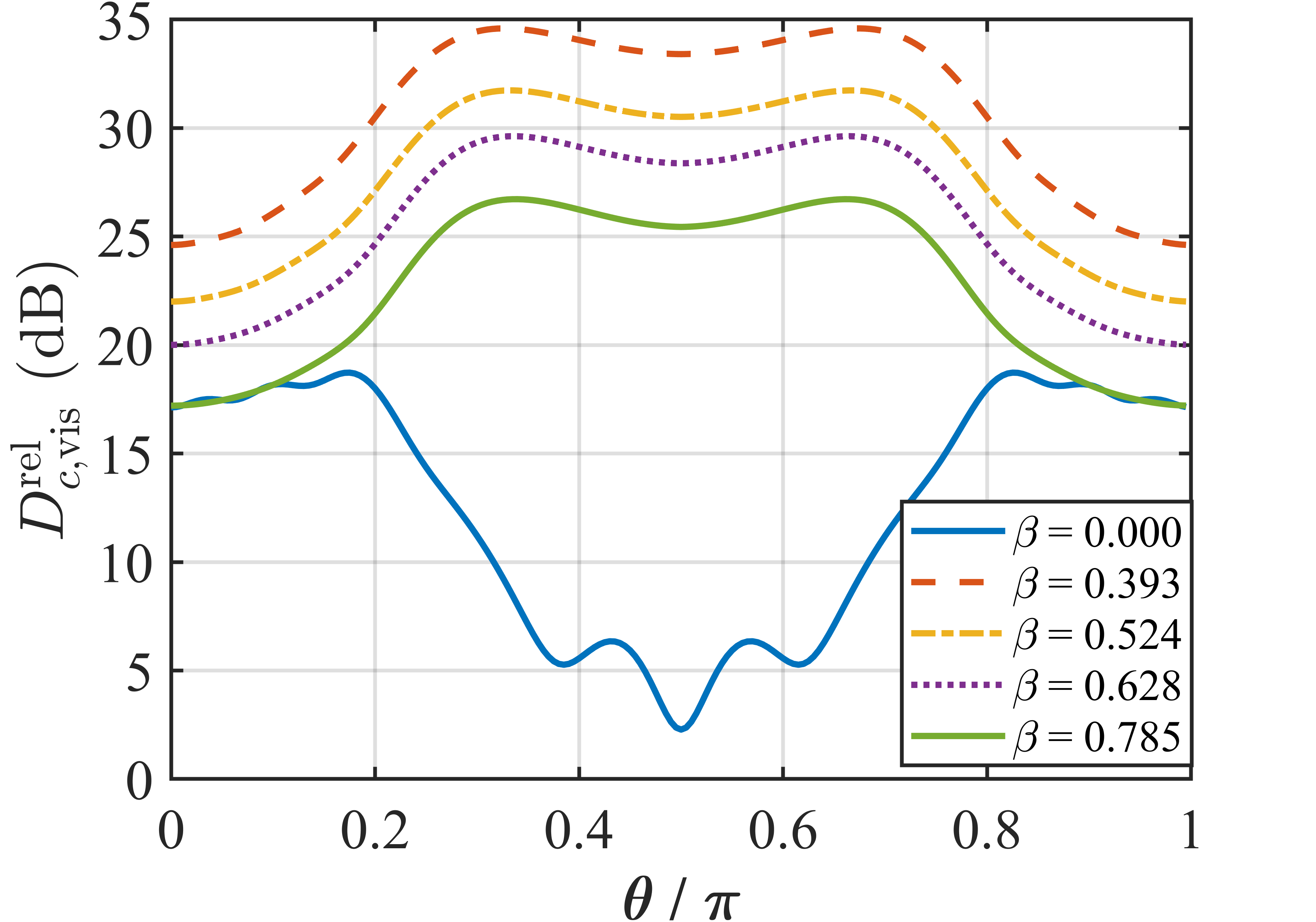}\label{fig:DirecFfpCM}}
\caption{Relative directivity on the endfire plane for different curvatures.}
\label{Fig:DirecFfp}
\end{figure}

Fig.~\ref{Fig:DirecFfp} shows the endfire plane ($\phi=0$, $\theta\in[0,\pi]$). Visibility changes the participating element set with angle. The nonmonotonic curvature trend reflects changes in visible aperture, projection, and phase response; MC further redistributes effective excitation.

\subsection{Spectral Efficiency}
TX and RX face $+y$ and $-y$, respectively. Their local normals are $\boldsymbol n_{T,n}=(0,\cos\psi_n,\sin\psi_n)^{\rm T}$ and $\boldsymbol n_{R,n}=(0,-\cos\psi_n,\sin\psi_n)^{\rm T}$. The MC-aware channel is \cite{8058474}
\begin{equation}
\mathbf H_c=\mathbf C_R\mathbf H_0\mathbf C_T,
\label{eq:Hc}
\end{equation}
where $\mathbf H_0$ is the uncoupled reference. To compare scattering environments \cite{11267247}, we use isotropic rays and a narrowband equivalent of CDL-A with one representative path per cluster. Retain directions with $\hat p_{T,y}\geq0$ and $\hat p_{R,y}\leq0$, and normalize their powers to $\sum_p P_p=1$. Each path has $g_p=\sqrt{P_p}e^{j\xi_p}$ with independent uniform phase $\xi_p$. Applying \eqref{eq:visibility} separately to its AoD and AoA gives
\begin{equation}
\widetilde{\mathbf H}_v=\sum_p g_p
(\mathbf V_{R,p}\boldsymbol a_{R,p})
(\mathbf V_{T,p}\boldsymbol a_{T,p})^{\rm H},
\label{eq:Hvisibility}
\end{equation}
where $\mathbf V_{T,p}$ and $\mathbf V_{R,p}$ are diagonal visibility masks and steering vectors have unit norm. Setting both masks to identity gives $\widetilde{\mathbf H}_0$.

\onecolumn

\twocolumn
\begin{figure*}[!t]
\centering
\begin{minipage}[t]{0.485\textwidth}
\centering
\includegraphics[width=0.90\linewidth]{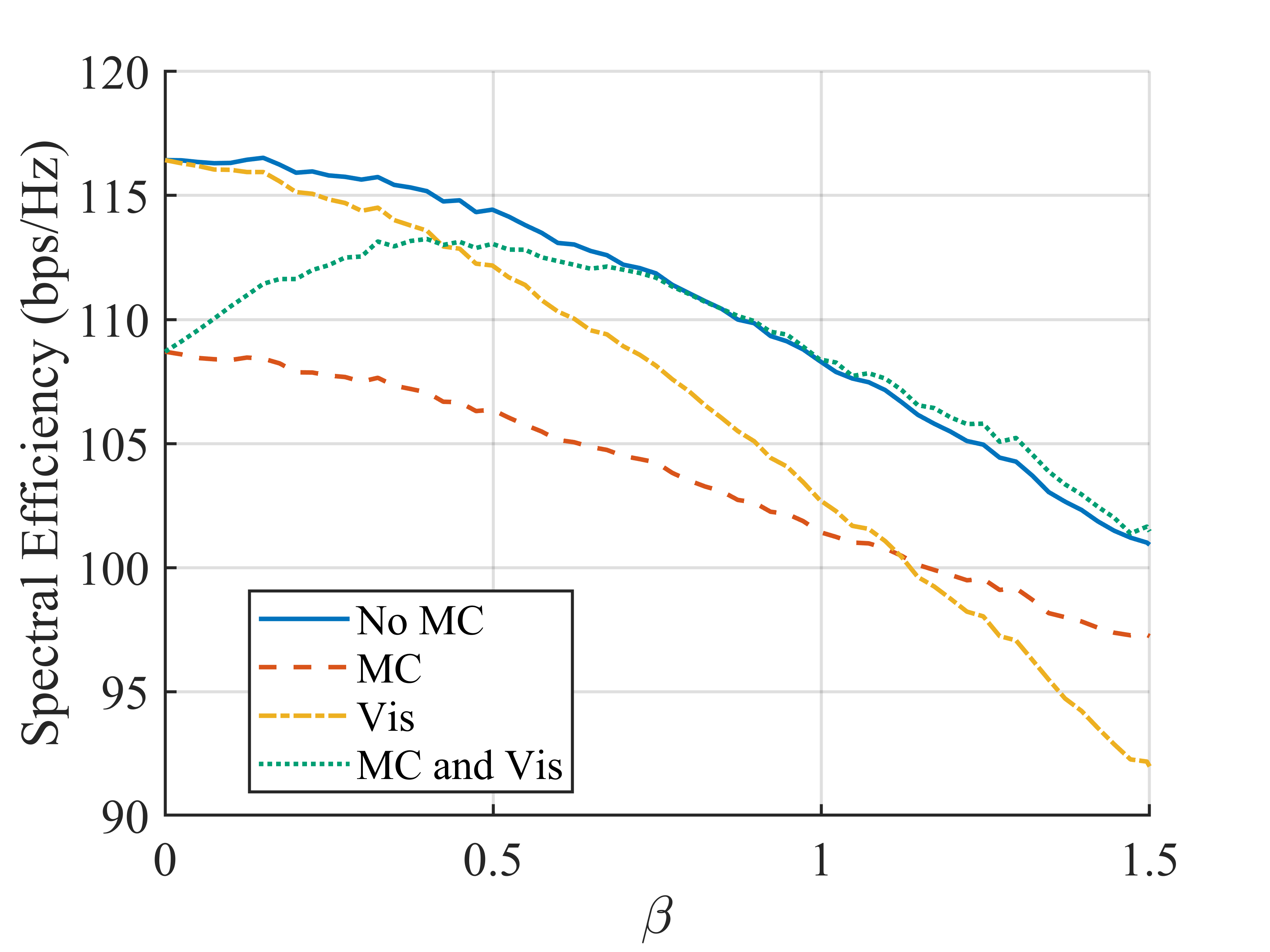}
\captionof{figure}{Spectral efficiency versus curvature under isotropic scattering.}
\label{Fig:CapaISO}
\end{minipage}\hfill
\begin{minipage}[t]{0.485\textwidth}
\centering
\includegraphics[width=0.90\linewidth]{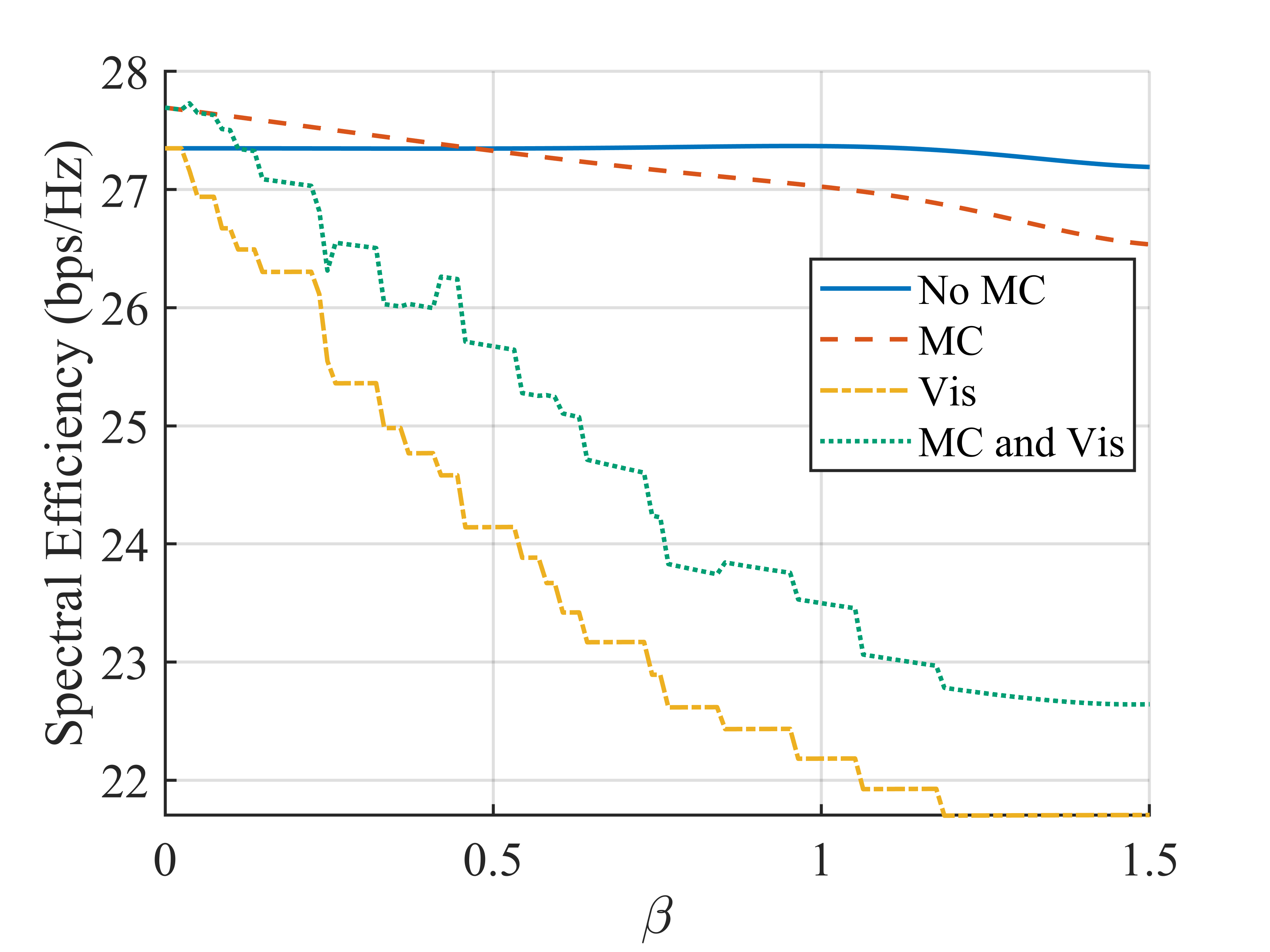}
\captionof{figure}{Spectral efficiency versus curvature under the CDL-A channel.}
\label{Fig:CapaCDLA}
\end{minipage}
\end{figure*}

\noindent For $N_T=N_R=MN$, set $c=\sqrt{N_TN_R}/\|\widetilde{\mathbf H}_0\|_F$ and use $\mathbf H_0=c\widetilde{\mathbf H}_0$ and $\mathbf H_v=c\widetilde{\mathbf H}_v$. Thus $\|\mathbf H_0\|_F^2=M^2N^2$; neither visibility nor MC is separately renormalized. The four cases use $\mathbf H_0$, $\mathbf C_R\mathbf H_0\mathbf C_T$, $\mathbf H_v$, and $\mathbf C_R\mathbf H_v\mathbf C_T$. With equal transmit-port power, perfect RX channel knowledge, and spatially white noise, spectral efficiency is
\begin{equation}
\mathrm{SE}=\mathbb E\!\left[\log_2\det\!\left(\mathbf I_{N_R}
+\frac{\rho}{N_T}\mathbf H_{\rm eff}\mathbf H_{\rm eff}^{\rm H}\right)\right],
\label{eq:SE}
\end{equation}
where $\rho=P_{\rm tot}/\sigma^2=10$ (10 dB), with total transmit power $P_{\rm tot}$ and noise variance $\sigma^2$ per RX port. The expectation uses 1000 realizations (Table~\ref{tab:channel_simulation_parameters}).

All four cases share the same path realization, while their common scaling retains visibility-induced power changes.

Under isotropic scattering (Fig.~\ref{Fig:CapaISO}), curvature lowers SE, and visibility causes additional degradation at large $\beta$. Without renormalization, the visibility channel retains power loss and changes in singular-value distribution; power and effective spatial DoF contributions are not separately quantified. MC redistributes amplitudes and phases, so the joint effect is not a simple sum of separate losses.

CDL-A (Fig.~\ref{Fig:CapaCDLA}) has lower SE, consistent with its more concentrated angular structure \cite{yan2026computational}. Without visibility, SE varies mildly with curvature; visibility produces a larger drop. Its step-like behavior follows from binary masks applied to discrete cluster directions. These results support accounting for both coupling and direction-dependent aperture participation.

\section{Conclusion}
We studied the directivity and spectral efficiency of a 2D HoloCuRA by jointly modeling curvature, correlation, MC, and visibility. Curvature changes the effective aperture; MC reshapes local fluctuations and channel responses, whereas visibility dominates global directivity and SE loss at large curvature. Both isotropic and CDL-A results support accounting for these effects in practical HMIMO systems.

\bibliographystyle{IEEEtran}  
\bibliography{references}

\end{document}